\documentclass[showpacs,preprintnumbers,amsmath,amssymb,prb]{revtex4}

\usepackage{graphicx}
\usepackage{bm}

\begin{document}

\title{Instability of the collinear phase in two-dimensional ferromagnet in strong in-plane magnetic field}

\author{A. V. Syromyatnikov}
 \email{syromyat@thd.pnpi.spb.ru}
\affiliation{Petersburg Nuclear Physics Institute, Gatchina, St.\ Petersburg 188300, Russia}

\date{\today}

\begin{abstract}

It is well-known that in thin ferromagnetic film with a net magnetization perpendicular to the film the collinear arrangement of spins is unstable in an in-plane field $H$ smaller than its saturation value $H_c$. Existence of a stripe phase was proposed with elongated domains of alternating direction of magnetization component perpendicular to the film. We consider in the present paper the strong-field regime $H<H_c$ and discuss the minimal microscopic model describing this phenomenon, two-dimensional Heisenberg ferromagnet with strong easy-axis anisotropy and dipolar forces. The noncollinear (stripe) phase is discussed using the technique of the Bose-Einstein condensation of magnons. Some previously unknown results are observed concerning the stripe phase. Evolution is established of the spin arrangement in the noncollinear phase upon the field rotation within the plane. We find a rapid decreasing of the period of the stripe structure as the field decreases. We demonstrate that spins components perpendicular to the film form a sinusoid in the noncollinear phase at $H\approx H_c$ that transforms to a step-like profile upon the field decreasing so that the domain wall density decreases from unity to a value much smaller than unity. The spin-wave spectrum in the noncollinear phase is discussed.

\end{abstract}


\maketitle

\section{Introduction}
\label{int}

\begin{figure}
\centering
\includegraphics[scale=0.1]{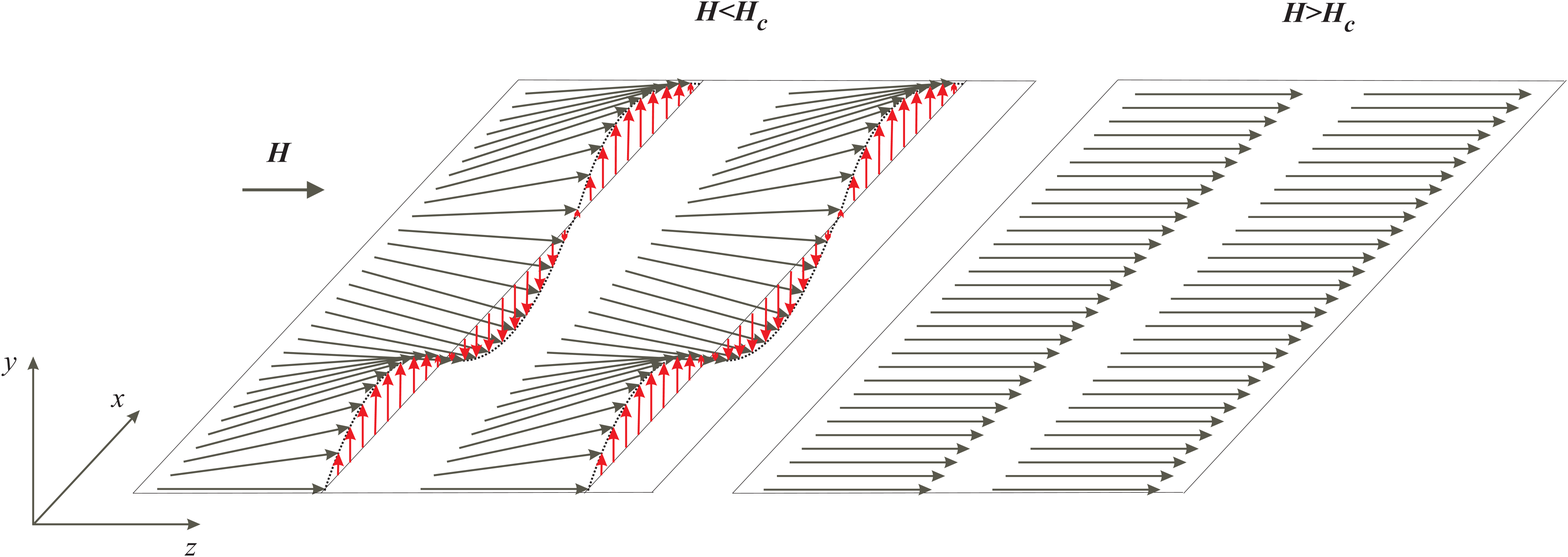}
\caption{(Color online). Two phases of the model (\ref{ham0}) in strong magnetic field $H$: those with collinear and noncollinear spin orderings at $H>H_c$ and $H<H_c$, respectively. Off-plane spins components are shown in red in the stripe phase (i.e., at $H<H_c$). It is obtained in the present paper that these components form a sinusoid at $H\approx H_c$ that transforms to a step-like profile upon the field decreasing as it is shown in Fig.~\ref{stripe}. Period of the stripe pattern rises as the field decreases as it is presented in Fig.~\ref{kcfig}.
\label{phases}} 
\end{figure}

Numerous fascinating magnetic properties of ultrathin films and (quasi-) two-dimensional (2D) magnetic materials attract much attention now that is stimulated also by the technological significance of such materials. \cite{rev,rev-jen} One of the most intriguing experimental findings was the discovery of the spin reorientation transition (SRT) in thin ferromagnetic films. \cite{pappas} Recent experimental and theoretical investigations have shown that SRT can be induced by the temperature, the film thickness or by an applied magnetic field. In most cases the rotation occurs during SRT from the perpendicular to an in-plane direction (or vice versa) of the film magnetization. In many materials with the net magnetization perpendicular to the film this reorientation is accompanied by stripe phases with elongated domains of alternating magnetization direction which are separated by domain walls with a finite width. Such transition was obtained first in Ref.~\cite{allen} in Co/Au(111) films. The reader is refered to comprehensive reviews \cite{rev,rev-jen} of numerous recent experimental and theoretical works in this field. It is well established now that the origin of SRT and stripe phases is in the subtle interplay between the short-range exchange, anisotropy and long-range dipolar interactions between spins. \cite{rev,rev-jen} Then, a realistic theoretical model of low-dimensional magnetic systems must include these three kinds of interactions so that the Hamiltonian has the form of the minimal microscopic model showing SRT and the stripe phase
\begin{eqnarray}
\label{ham0}
{\cal H} &=& -\frac12 \sum_{l\ne m} \left(J_{lm}\delta_{\rho\beta} + Q_{lm}^{\rho\beta}\right) S_l^\rho  S_m^\beta 
-A\sum_l \left(S_l^y\right)^2 - H \sum_l S_l^z,\\
\label{q}
Q_{lm}^{\rho\beta} &=& (g\mu)^2\frac{3R_{lm}^\rho R_{lm}^\beta - \delta_{\rho\beta}R_{lm}^2}{R_{lm}^5},
\end{eqnarray}
where we direct $y$ axis perpendicularly to the lattice as it is shown in Fig.~\ref{phases} and the last term describes the Zeeman energy in the in-plane field discussed below.

It is demonstrated in Ref.~\cite{stripes} that at $H=0$ and $T=0$ the stripe arrangement of spins in the model \eqref{ham0} has slightly smaller energy than any collinear arrangement if $2A>\alpha\omega_0$ (i.e., if the easy direction is normal to the film), where
\begin{equation}
\label{alpha}
\alpha = \frac{3v_0}{8\pi}\sum_i \frac{1}{R_i^3}
\end{equation}
is a constant that is equal approximately to $1.078$ for the simple square lattice,
\begin{equation}
\label{o0}
\omega_0 = 4\pi (g\mu)^2
\end{equation}
is the characteristic dipolar energy and we set the lattice spacing to be equal to unity. Meantime it was found in Ref.~\cite{stripes} that the period of the stripe structure rises rapidly as the value of the anisotropy increases so that the domain width is macroscopically large (i.e., larger than $10^8$ lattice spacing) for $2A>1.4\alpha\omega_0$. It is implied hereafter that $J\gg\omega_0$, as it usually is. We assume in the present paper that $2A>1.4\alpha\omega_0$ and consider macroscopically large unidomain sample with all spins directed perpendicularly to the film at $H=0$. It is assumed below also that $J\gg A$.

Existence of a stripe phase in this case was pointed out long time ago \cite{emills} in an interval of in-plane magnetic field $H_c^\prime<H<H_c$, where $H_c\propto A$ is the saturation field,
\begin{equation}
\label{hc1sw}
H_c^\prime=H_c-\frac32Dk_{c0}^2,
\end{equation}
$D$ is the spin-wave stiffness that is equal, in particular, to $SJ$ for simple square lattice with exchange coupling constant $J$ between neighboring spins, and 
\begin{equation}
\label{kc}
k_{c0} = \frac{S\omega_0}{4D}.
\end{equation}
The experimental evidence for the in-plane field induced collinear phase instability has been reported recently in Ni(001) ultrathin films at room temperature. \cite{exp_str} It was pointed out in Ref.~\cite{emills} that the spin-wave spectrum calculated in the linear spin-wave approximation assuming a collinear spin arrangement is unstable when $H_c^\prime<H<H_c$. Importantly, the instability takes place not at zero momentum but at a finite incommensurate one ${\bf k}_{c0}$ which direction is not fixed in the linear spin-wave approximation and which value is given by Eq.~\eqref{kc}. As a result it was proposed that the model \eqref{ham0} shows three phases in a nonzero in-plane field, collinear one at $H>H_c$, collinear canted phase at $H<H_c^\prime$, and the noncollinear one appearing at $H_c^\prime<H<H_c$. The properties of the noncollinear phase have not been clarified yet. It was proposed in Ref.~\cite{emills} that there should be a domain pattern at $H_c^\prime<H<H_c$ with the period equal to $2\pi/k_{c0}$ forming by spins components perpendicular to the film similar to that discussed in Ref.~\cite{stripes} for $H=0$. 

It should be pointed out however that the stability of the spin-wave spectrum is not a criterion for the corresponding state to be the ground state. Then, it is not clear at which field $H_c^\prime$ the transition takes place from the stripe phase to a canted collinear one. It is not clear also whether such a transition happens at all because the stripe phase has lower energy at $H=0$, as it is shown in Ref.~\cite{stripes}.

Despite great recent interest to exotic non-collinear phases, that of the model \eqref{ham0} in strong-field regime has not been discussed thoroughly yet. The aim of the present paper is to fill up somewhat this gap. We discuss first $1/S$ corrections to the spectrum in the collinear phase at $H>H_c$ and show that they fix the direction of ${\bf k}_{c0}$ at which the spectrum instability takes place and which absolute value is given by Eq.~\eqref{kc}. We find the dependence of ${\bf k}_{c0}$ direction on the field direction within the plane.

The noncollinear phase is discussed below using the technique of the Bose-Einstein condensation of magnons suggested in Ref.~\cite{bat84}. The transition to the stripe phase at $H=H_c$ corresponds within this technique to a "condensation" of magnons at states characterized by a momentum ${\bf k}_c\|{\bf k}_{c0}$ and its harmonics (i.e., $n{\bf k}_c$, where $n$ is integer). It is the unusual result of the present analysis that the vector ${\bf k}_c$ varies with the field (remaining parallel to ${\bf k}_{c0}$) so that $k_c=k_{c0}$ at $H=H_c$ and $k_c<k_{c0}$ at $H<H_c$. This unusual behavior is related to condensation of magnons at states characterized by harmonics of ${\bf k}_c$. Remember that the condensation in antiferromagnetic \cite{bat84} and ferromagnetic \cite{ibec} systems takes place at antiferromagnetic and zero (commensurate) vectors, respectively, which do not depend on magnetic field. Thus, we find that the period of the stripe structure given by $2\pi/k_c$ rises very fast as the field decreases.

We demonstrate that the energy of the stripe phase is lower than the energy of any collinear spin arrangement at $H<H_c$ but the difference between these energies decreases as the field reduces. 

We find values of spins components perpendicular to the film forming the domain pattern. It is shown that spins components perpendicular to the film form a sinusoid in the noncollinear phase at $H\approx H_c$ that transforms to a step-like profile upon the field decreasing so that the domain wall density decreases from unity to a value much smaller than unity.

We discuss below also the spin-wave spectrum in the noncollinear phase.

The rest of the paper is organized as follows. We discuss the Hamiltonian transformation and technique in Sec.\ref{hamtr}. Magnon spectrum is discussed in the linear spin-wave approximation in Sec.~\ref{ms}. First $1/S$ corrections to the spectrum at $H>H_c$ is discussed in Sec.~\ref{specren}. Properties of the noncollinear phase are considered in Sec.~\ref{noncol}. Sec.~\ref{con} contains our conclusion. One appendix is added with some details of calculations.

\section{Hamiltonian transformation and technique}
\label{hamtr}

It is implied below that the in-plane magnetic field is directed arbitrary relative to the lattice. Taking the Fourier transformation we have from Eq.~(\ref{ham0})
\begin{equation}
\label{ham}
{\cal H} = -\frac12 \sum_{\bf k}\left(J_{\bf k}\delta_{\rho\beta} + Q_{\bf k}^{\rho\beta}\right) S_{\bf k}^\rho  S_{-\bf k}^\beta
-A\sum_{\bf k} S_{\bf k}^y S_{-\bf k}^y
-H \sqrt{\mathfrak N} S_{\bf 0}^z,
\end{equation}
where $J_{\bf k} = \sum_l J_{lm}\exp(i{\bf k R}_{lm})$, $Q_{\bf k}^{\rho\beta} = \sum_l Q_{lm}^{\rho\beta}\exp(i{\bf k R}_{lm})$ and $\mathfrak N$ is the number of spins in the lattice. Dipolar tensor $Q_{\bf k}^{\rho\beta}$ possesses the well-known properties \cite{mal,rev} at $k\ll1$, which are independent of the lattice type and the orientation of $x$ and $z$ axes relative to the lattice,
\begin{eqnarray}
\label{qsmall}
Q_{\bf k}^{\rho\beta} &=& \omega_0\left( \frac\alpha3 \delta_{\rho\beta} - \frac k2 \frac{k_\rho k_\beta}{k^2} \right), \mbox{ where } \rho,\beta=x,z,\\
\label{qy}
Q_{\bf k}^{y\beta} &=& \omega_0 \left( -\frac23 \alpha + \frac k2 \right)\delta_{y\beta},
\mbox{ where } \beta=x,y,z,
\end{eqnarray}
where $\alpha$ and $\omega_0$ are given by Eqs.~\eqref{alpha} and \eqref{o0}, respectively. It is seen from Eqs.~(\ref{ham0}), (\ref{qsmall}) and (\ref{qy}) that dipolar forces lead to easy-plane anisotropy in the energy of the classical 2D FM with $y$ to be a hard axis. \cite{mal} Quantum and thermal fluctuations lead also to an in-plain anisotropy (order-by-disorder effect). \cite{anis_dan,i2d,i}

We use in the present paper the well-known representation of spins components via Bose-operators $a$ and $a^\dagger$: \cite{sw}
\begin{equation}
\label{hpls}
S_i^- \approx \sqrt{2S} a_i^\dagger \left(1 - \frac{a_i^\dagger a_i}{4S}\right),
\qquad 
S_i^\dagger \approx \sqrt{2S} \left(1 - \frac{a_i^\dagger a_i}{4S}\right) a_i,
\qquad  
S_i^z = S - a_i^\dagger a_i
\end{equation}
which can be obtained from the Holstein-Primakoff transformation \cite{hp} by expanding the square roots up to the first terms. 

To perform the calculations it is convenient to introduce the following retarded Green's functions: $G(\omega,{\bf k}) = \langle a_{\bf k}, a^\dagger_{\bf k} \rangle_\omega$, $F(\omega,{\bf k}) = \langle a_{\bf k}, a_{-\bf k} \rangle_\omega$, ${\overline G}(\omega,{\bf k}) = \langle a^\dagger_{-\bf k}, a_{-\bf k} \rangle_\omega = G^*(-\omega,-{\bf k})$ and $F^\dagger (\omega,{\bf k}) = \langle a^\dagger_{-\bf k}, a^\dagger_{\bf k} \rangle_\omega = F^*(-\omega,-{\bf k})$. We have the following set of Dyson equations:
\begin{equation}
\label{eqfunc}
\begin{array}{l}
G(\omega,{\bf k}) = G^{(0)}(\omega,{\bf k}) + G^{(0)}(\omega,{\bf k}){\overline \Sigma}(\omega,{\bf k})G(\omega,{\bf k}) + G^{(0)}(\omega,{\bf k}) [B_{\bf k} + \Pi(\omega,{\bf k})] F^\dagger(\omega,{\bf k}),\\
F^\dagger(\omega,{\bf k}) = {\overline G}^{(0)}(\omega,{\bf k}) \Sigma(\omega,{\bf k})F^\dagger(\omega,{\bf k}) + {\overline G}^{(0)}(\omega,{\bf k}) [B_{\bf k} + \Pi^\dagger(\omega,{\bf k}) ]G(\omega,{\bf k}),
\end{array}
\end{equation}
where $G^{(0)}(\omega,{\bf k}) = (\omega - E_{\bf k}+i\delta)^{-1}$ is the bare Green's function, 
$
\Sigma(\omega,{\bf k}) = \overline \Sigma(-\omega,-{\bf k})^*, 
$
$
\Pi(\omega,{\bf k}) = \Pi^\dagger(-\omega,-{\bf k})^*
$ 
are the self-energy parts and $E_{\bf k}$ and $B_{\bf k}$ are coefficients in the bilinear part of the Hamiltonian
\begin{equation}
\label{h2}
{\cal H}_2 = \sum_{\bf k} \left[E_{\bf k} a^\dagger_{\bf k}a_{\bf k} + \frac{B_{\bf k}}{2} \left(a_{\bf k}a_{-\bf k} + 
a^\dagger_{\bf k}a^\dagger_{-\bf k}\right)\right]
\end{equation}
which are presented below. Solving Eqs.~(\ref{eqfunc}) one obtains
\begin{eqnarray}
G(\omega,{\bf k}) &=& \frac{\omega + E_{\bf k} + \Sigma(\omega,{\bf k})}{{\cal D}(\omega,{\bf k})},\nonumber\\
\label{gf}
F(\omega,{\bf k}) &=& -\frac{B_{\bf k} + \Pi(\omega,{\bf k})}{{\cal D}(\omega,{\bf k})},
\end{eqnarray}
where
\begin{eqnarray}
\label{d}
{\cal D}(\omega,{\bf k}) &=& (\omega+i\delta)^2 - \epsilon_{\bf k}^2 - \Omega(\omega,{\bf k}),\\
\label{spec0}
\epsilon_{\bf k}^2 &=& E_{\bf k}^2 - B_{\bf k}^2,\\
\label{o}
\Omega(\omega,{\bf k}) &=& E_{\bf k}(\Sigma + \overline{\Sigma}) - B_{\bf k}(\Pi + \Pi^\dagger) - (\omega + i\delta)(\Sigma - \overline{\Sigma}) - \Pi\Pi^\dagger + \Sigma \overline{\Sigma},
\end{eqnarray}
and $\epsilon_{\bf k}$ is the spin-wave spectrum in the linear spin-wave approximation. Quantity $\Omega(\omega,{\bf k})$ given by Eq.~(\ref{o}) describes renormalization of the spin-wave spectrum square. We find $\Omega(\omega,{\bf k})$ within the first order of $1/S$ in Sec.~\ref{specren}.

\section{Classical magnon spectra}
\label{ms}

\subsection{$H>H_c$}

Let us assume first that the field is so strong that all spins lie within the plane (see Fig.~\ref{phases}). After substitution of Eqs.~(\ref{hpls}) into Eq.~(\ref{ham}) the Hamiltonian has the form ${\cal H} = E_0 + \sum_{i=1}^6 {\cal H}_i$, where $E_0$ is the ground state energy and ${\cal H}_i$ denote terms containing products of $i$ operators $a$ and $a^\dagger$. ${\cal H}_1=0$ because it contains only $Q_{\bf 0}^{\rho\beta}$ with $\rho\ne\beta$. For ${\cal H}_2$ one has Eq.~(\ref{h2}), where
\begin{eqnarray}
\label{els}
E_{\bf k} &=& S(J_{\bf 0} - J_{\bf k}) - \frac S2\left( Q_{\bf k}^{xx} + Q_{\bf k}^{yy} - \frac{2\omega_0\alpha}{3} \right) + H - AS\left(1-\frac{1}{2S}\right) \nonumber\\
& \stackrel{k\ll1}{\approx} & Dk^2 - \frac{S\omega_0}{4}k\cos^2\phi_{\bf k} + H - AS\left(1-\frac{1}{2S}\right) + \frac{S\omega_0\alpha}{2},
\\
\label{bls}
B_{\bf k} &=& \frac S2 \left( Q_{\bf k}^{yy} - Q_{\bf k}^{xx} \right) + AS\left(1-\frac{1}{4S}\right)
\stackrel{k\ll1}{\approx} -\frac{S\omega_0\alpha}{2} + \frac{S\omega_0}{4}k (1 + \sin^2\phi_{\bf k})\nonumber\\
&&{} + AS\left(1-\frac{1}{4S}\right),
\end{eqnarray}
where $\phi_{\bf k}$ is the angle between $\bf k$ and the magnetization and the expressions after $\stackrel{k\ll1}{\approx}$ are approximate values of the corresponding quantities at $k\ll1$. We find for the square of the magnon spectrum from Eqs.~(\ref{spec0}), (\ref{els}) and (\ref{bls}) in accordance with the well-known result \cite{emills}
\begin{eqnarray}
\label{spec1}
\epsilon_{\bf k}^2 &\stackrel{k\ll1}{\approx}& \left(Dk^2 + \frac{S\omega_0}{2}k\sin^2\phi_{\bf k} + H \right)\left(D(k-k_{c0})^2 + \Lambda^{(0)} \right) - \frac A2B_{\bf k},\\
\label{lambda}
\Lambda^{(0)} &=& H - H_c^{(0)},\\
\label{hc20}
H_c^{(0)} &=& 2SA\left(1-\frac{1}{2S}\right) - S\omega_0\alpha + Dk_{c0}^2,
\end{eqnarray}
where $H_c^{(0)}$ is the classical value of the saturation field and $k_{c0}$ is given by Eq.~\eqref{kc}. The second term in the first brackets in Eq.~\eqref{spec1} is negligible in the considered case of $A\gg\omega_0$. The last term in Eq.~(\ref{spec1}) is formally of the next order in $1/S$ compared to the first one. We show below that corrections from the Hartree-Fock diagram presented in Fig.~\ref{diag}(a) cancel it and leads to renormalization of the critical field $H_c^{(0)}$. We point out also that the anisotropy constant $A$ should be accompanied by the factor $1-1/2S$ in all expressions for observable quantities because there is no one-ion anisotropy for $S=1/2$. \cite{chub} It is seen that in the zeroth order in $1/S$ (the last term in Eq.~(\ref{spec1}) should be discarded) the second bracket in Eq.~(\ref{spec1}) is positive at $H\ge H_c^{(0)}$ and the spectrum is stable. On the other hand at $H<H_c^{(0)}$ the instability of the spectrum arises at momentum ${\bf k}_{c0}$. Notice that the direction of ${\bf k}_{c0}$ is not fixed within the spin-wave approximation. Meantime we show below that there are anisotropic corrections to the spectrum of the first order in $1/S$, which fix the direction of the vector ${\bf k}_{c0}$ at which the instability arises leaving, however, the equivalence between ${\bf k}_{c0}$ and $-{\bf k}_{c0}$.

\subsection{$H<H_c$}

Simple calculation leads to the following result for the spectrum at $k\ll1$ assuming the collinear canted spin arrangement at $H<H_c$:
\begin{equation}
\label{speca}
\epsilon_{\bf k}^2 \approx \left(Dk^2 + 2SA\left(1-\frac{1}{2S}\right) \right)\left(D(k-k_{c0})^2 + 2(H_c^\prime - H) \right) - S^2(Q^{xz}_{\bf k})^2\sin^2\theta,
\end{equation}
where $k_{c0}$ and $H_c^\prime$ are given by Eqs.~\eqref{kc} and \eqref{hc1sw}, respectively, $\cos\theta=H/(2SA(1-1/2S)-S\omega_0\alpha)$, $\theta$ is the canting angle between spins and the field, and we have discarded terms proportional to $A$ similar to the last one in Eq.~\eqref{spec1} which are of the next order in $1/S$. The second bracket in Eq.~\eqref{speca} vanishes at $k=k_{c0}$ when $H=H_c^\prime$. The last term in Eq.~\eqref{speca} gives negligibly small correction to $H_c^\prime$. The spectrum \eqref{speca} is unstable at $H>H_c^\prime$. This result was obtained first in Ref.~\cite{emills}. It was proposed there that a noncollinear phase arises at $T=0$ in the interval $H_c^\prime<H<H_c$, where $H_c$ and $H_c^\prime$ are given up to quantum $1/S$ corrections by Eqs.~\eqref{hc20} and \eqref{hc1sw}, respectively. Meantime we show below that the energy of the collinear phase at $H=H_c^\prime$ is higher than that of the noncollinear one and the value of the critical field $H_c^\prime$ is smaller (if it is nonzero at all) than that given by Eq.~\eqref{hc1sw} and derived from the spin-wave analysis. 

\section{Spectrum renormalization at $H>H_c$}
\label{specren}

\begin{figure}
\centering
\includegraphics[scale=0.7]{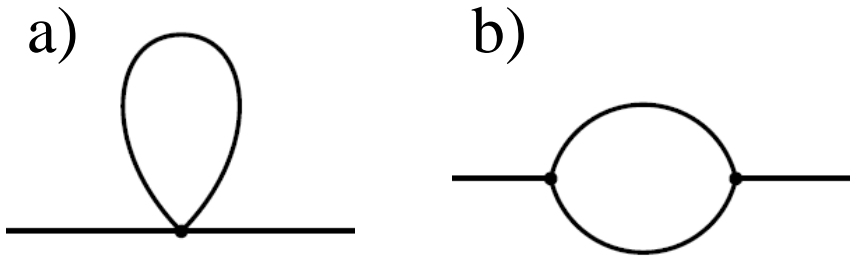}
\caption{Diagrams of the first order in $1/S$ for self-energy parts. Diagram (a) stems from four-magnon terms in the Hamiltonian whereas (b) comes from three-magnon terms.
\label{diag}} 
\end{figure}

Corrections of the first order in $1/S$ to the spectrum stem from diagrams shown in Fig.~\ref{diag}. To calculate them one should take into account terms ${\cal H}_3$ and ${\cal H}_4$ in the Hamiltonian which have the form
\begin{eqnarray}
\label{h3}
{\cal H}_3 &=& \sqrt{\frac {S}{2 \mathfrak N}} \sum_{{\bf k}_1 + {\bf k}_2 + {\bf k}_3 = {\bf 0}} 
Q_2^{xz} a^\dagger_{-1}\left(a^\dagger_{-2} + a_2\right)a_3,\\
{\cal H}_4 &=& \frac{1}{8\mathfrak N}\sum_{{\bf k}_1 + {\bf k}_2 + {\bf k}_3 + {\bf k}_4 = {\bf 0}} 
\left(
\left[4A + 2(J_1 + J_3 - 2J_{1+3}) - (Q_1^{zz} + Q_3^{zz} + 4Q_{1+3}^{zz})\right]
a^\dagger_{-1}a^\dagger_{-2}a_3a_4 \right.\nonumber\\
\label{h4ls}
&&{}\left.
+ 
\left[
Q_2^{xx}-Q_2^{yy} - 2A
\right]
a^\dagger_{-1} 
\left[
a_2a_3 + a^\dagger_{-2} a^\dagger_{-3}
\right] 
a_4 
\right),
\end{eqnarray}
where we drop index $\bf k$ in Eqs.~(\ref{h3}) and (\ref{h4ls}).

We have found in accordance with the conclusion of Ref.~\cite{emills} that the loop diagram presented in Fig.~\ref{diag}(b) is much smaller than the Hartree-Fock one shown in Fig.~\ref{diag}(a). One obtains for the contribution to $\Omega(\omega,{\bf k})$ from the Hartree-Fock diagram
\begin{subequations}
\label{o4}
\begin{eqnarray}
\Omega(\omega,{\bf k}) &=& 
B_{\bf k} \frac{1}{2S\mathfrak N} \sum_{\bf q} B_{\bf q} \\
&&{} -
\epsilon_{\bf k}^2 \frac{1}{\mathfrak N} \sum_{\bf q} \frac{E_{\bf q} - \epsilon_{\bf q}}{S\epsilon_{\bf q}}\\
&& {}+
\frac{1}{\mathfrak N}\sum_{\bf q} \frac{1}{\epsilon_{\bf q}} 
\left(E_{\bf k}E_{\bf q} + B_{\bf k}B_{\bf q}\right) 
\left(J_{\bf q} - J_{\bf k+q} + Q^{zz}_{\bf q} - Q^{zz}_{\bf k+q}\right)\\
&& {} + E_{\bf k} \frac{1}{\mathfrak N}\sum_{\bf q} 
\left[ 
\frac{E_{\bf q} - \epsilon_{\bf q}}{\epsilon_{\bf q}}\left(\frac HS +A - \frac32 Q_{\bf q}^{zz}\right)
+
\frac{B_{\bf q}^2}{S\epsilon_{\bf q}}
\right]\\
&& {} + B_{\bf k} \frac{1}{\mathfrak N}\sum_{\bf q} 
\left[ 
\frac{E_{\bf q} - \epsilon_{\bf q}}{2S\epsilon_{\bf q}}B_{\bf q}
+
\frac{B_{\bf q}}{\epsilon_{\bf q}} 
\left(
\frac{E_{\bf q} - \epsilon_{\bf q}}{2S} + A - \frac32 Q_{\bf q}^{zz}
\right)
\right].
\end{eqnarray}
\end{subequations}
Term (a) in Eq.~(\ref{o4}) is equal in the first order in $1/S$ to $B_{\bf k} A/2$ and it cancels the last term in Eq.~(\ref{spec1}). Term (\ref{o4}b) does not renormalize the bare spectrum in the first order in $1/S$. Term (\ref{o4}c) gives rise to the anisotropic corrections to the spectrum. 

It is convenient to use evident relations $E_{\bf k} = (E_{\bf k} + B_{\bf k})/2 + (E_{\bf k} - B_{\bf k})/2$ and $B_{\bf k} = (E_{\bf k} + B_{\bf k})/2 - (E_{\bf k} - B_{\bf k})/2$ to extract combinations proportional to $E_{\bf k} - B_{\bf k}$ and $E_{\bf k} + B_{\bf k}$ from terms (c)--(e) of Eq.~(\ref{o4}). Terms proportional to $E_{\bf k} - B_{\bf k}$ and $E_{\bf k} + B_{\bf k}$ renormalize the first and the second brackets in Eq.~(\ref{spec1}), respectively. As we are interested in the phase transition to the noncollinear phase, we focus on renormalization of the second bracket in Eq.~(\ref{spec1}) and consider only terms proportional to $E_{\bf k} + B_{\bf k}$.

\begin{figure}
\centering
\includegraphics[scale=0.7]{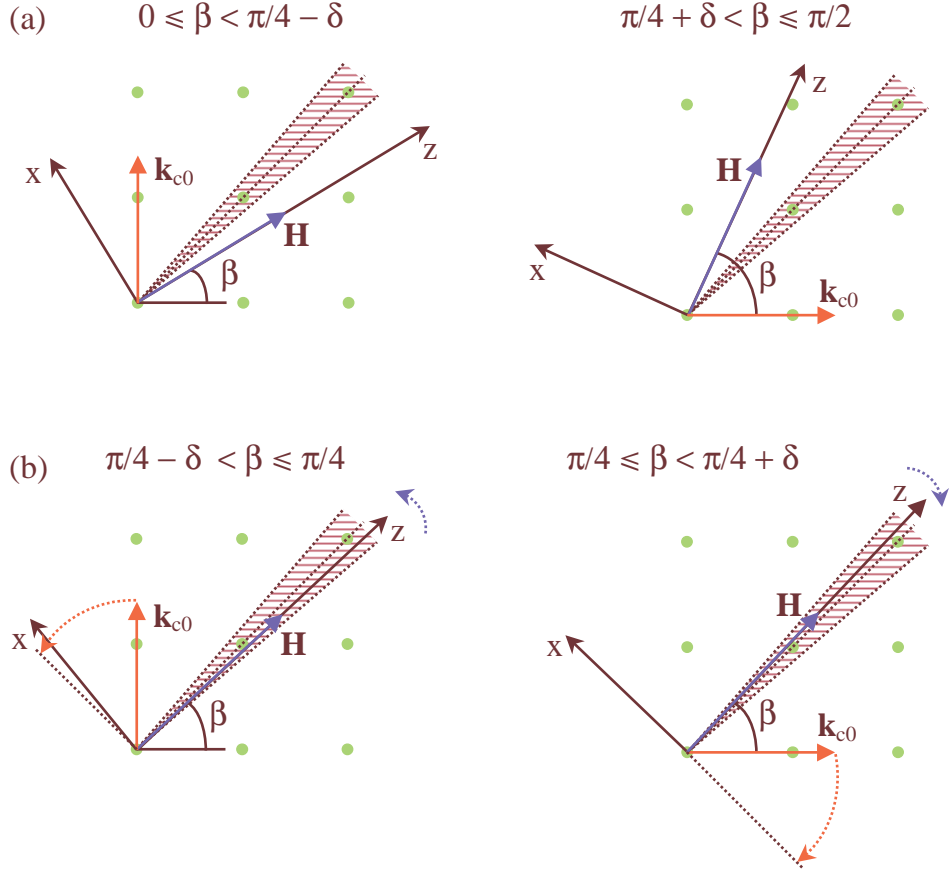}
\caption{(Color online). It is assumed in this paper that the field $\bf H$ lies within the plane of the lattice and is directed by an angle $\beta$ to the square edge. This figure shows the direction of the vector ${\bf k}_{c0}$ at which instability arises of the classical spin-wave spectrum at $H<H_c$. Domains in the stripe phase are elongated perpendicularly to ${\bf k}_{c0}$. Slide (a) is for the case when $\beta$ lays outside the narrow interval $(\pi/4-\delta,\pi/4+\delta)$ (cross-hatched area in the figure), where $\delta\ll1$ is given by Eq.~(\ref{delta}). Slide (b) illustrates the rotation of ${\bf k}_{c0}$ when the field rotates inside the interval $(\pi/4-\delta,\pi/4+\delta)$. The direction of the corresponding rotations are indicated by dashed arrows. We have ${\bf k}_{c0}\perp\bf H$ when $\beta=\pi/4$. Notice that ${\bf k}_{c0}$ and $-{\bf k}_{c0}$ are equivalent.
\label{angle}} 
\end{figure}

We assume for definiteness in the particular calculations below that $J\gg A\gg\omega_0$. As a result of simple but tedious calculations one has from Eq.~(\ref{o4}) in the leading order of $\omega_0/J$ and $A/J$ for the important for us terms
\begin{eqnarray}
\label{o4t0}
\Omega(\omega,{\bf k}) &\sim&
\frac A2B_{\bf k} \nonumber\\
&&{}+ 
(E_{\bf k} + B_{\bf k}) 
\left(
k^2\omega_0
[
X\cos(2\phi_{\bf k}+2\beta)\cos2\beta + Y \cos^2\phi_{\bf k}
]
+
\frac{A}{\mathfrak N}\sum_{\bf q}
\frac{B_{\bf q}}{\epsilon_{\bf q}}
\right),\\
\label{x}
X &=& \frac{D}{16S\omega_0 \mathfrak N} \sum_{\bf q}\frac{Q_{\bf q}^{xx} - Q_{\bf q}^{zz}}{J_{\bf 0} - J_{\bf q}}
(\cos q_x - \cos q_z),\\
\label{y}
Y &=& \frac{3}{2^7\pi} \sqrt{\frac{2AS}{D}}\ln\left(32\frac{ASD}{(S\omega_0)^2}\right),
\end{eqnarray}
where $\beta$ is the angle between magnetic field and an edge of the square (see Fig.~\ref{angle}). The constant $X$ should be calculated numerically because summation over large momenta is important in Eq.~(\ref{x}) and one cannot use Eqs.~(\ref{qsmall}) and (\ref{qy}) for the dipolar tensor components. This calculation can be carried out using the dipolar sums computation technique (see, e.g., Ref.~\cite{sums} and references therein) with the result $X\approx0.0087$ for exchange coupling between only nearest neighbor spins on the simple square lattice. 

Terms in Eq.~(\ref{o4t0}) proportional to $X$ and $Y$ stem from parts of the term (c) in Eq.~(\ref{o4}) proportional to $J_{\bf q} - J_{\bf k+q}$ and $Q^{zz}_{\bf q} - Q^{zz}_{\bf k+q}$, respectively. They make negligibly small correction to the constant $D$ in the second bracket in Eq.~(\ref{spec1}). Meantime they are very important for our consideration because they fix the direction of the momentum at which the spectrum instability arises: ${\bf k}_{c0}$ is directed so that this term to be minimal at $\phi_{\bf k}=\phi_{{\bf k}_{c0}}$. Notice that anisotropic terms are invariant according to replacement $\phi_{\bf k}\mapsto\phi_{\bf k}\pm\pi$. Then ${\bf k}_{c0}$ and $-{\bf k}_{c0}$ are equivalent. One obtains after simple calculations that the anisotropic part in Eq.~(\ref{o4t0}) has a minimum at $\phi_{{\bf k}_{c0}}$ given by
\begin{equation}
\label{pkcs}
\tan 2\phi_{{\bf k}_{c0}} = -\frac{\sin4\beta}{Y/X + 2\cos^22\beta},
\end{equation}
where the solution should be taken with $\cos 2\phi_{{\bf k}_{c0}}<0$. The term $Y/X$ plays in Eq.~\eqref{pkcs} only when $\beta\approx\pi/4$ because $X\gg Y$. Simple calculation shows that the first and the second terms in denominator in Eq.~\eqref{pkcs} are equal at $\beta=\pi/4\pm\delta$, where
\begin{equation}
\label{delta}
\delta = \sqrt{\frac{Y}{8X}} \sim \left(\frac{AS}{D}\right)^{1/4} \ll 1.
\end{equation}
Then, when $\beta$ lays outside the narrow interval $(\pi/4-\delta,\pi/4+\delta)$ the term with $Y$ can be discarded in Eq.~(\ref{pkcs}) and one has for $\phi_{{\bf k}_{c0}}$
\begin{equation}
\label{pkcl}
\phi_{{\bf k}_{c0}} = 
\left\{
\begin{array}{ll}
\pi/2-\beta, & \mbox{ if } \beta\in[0,\pi/4-\delta), \\
-\beta, &  \mbox{ if } \beta\in(\pi/4+\delta,\pi/2].
\end{array}
\right.
\end{equation}
Eq.~(\ref{pkcl}) signifies that ${\bf k}_{c0}$ does not rotate with the field and is directed along the square edge as it is shown in Fig.~\ref{angle}(a). On the other hand, the term with $Y$ comes into play when $\beta$ lays in the interval $(\pi/4-\delta,\pi/4+\delta)$ and ${\bf k}_{c0}$ rotates in the same direction as the field $\bf H$ does so that ${\bf k}_{c0}$ turns out to be perpendicular to $\bf H$ when $\beta=\pi/4$ (see Fig.~\ref{angle}(b)).

The last term in Eq.~(\ref{o4t0}) leads to quantum renormalization of the critical field $H_c$ which has the form
\begin{equation}
\label{hc2t0}
H_c = H_c^{(0)} - \frac{A^2S}{4\pi D} \ln\left(\frac{D}{2AS}\right)
\end{equation}
when $\ln(D/AS)\gg1$, where $H_c^{(0)}$ is given by Eq.~(\ref{hc20}). It is seen from Eq.~(\ref{hc2t0}) that quantum fluctuations reduce the value of the saturation field. As a result we have for the spin-wave spectrum at $k\ll1$
\begin{eqnarray}
\label{specres}
\epsilon_{\bf k}^2 &\approx& \left(Dk^2 + H \right)\left(D(k-k_{c0})^2 + \Lambda + \delta(\phi_{\bf k})\right),\\
\label{lambdares}
\Lambda &=& H - H_c,
\end{eqnarray}
where $H_c$ is given by Eq.~\eqref{hc2t0} and $\delta(\phi_{\bf k})$ stands for the small anisotropic terms discussed above.

\section{The noncollinear phase}
\label{noncol}

We discuss in this section properties of the noncollinear phase. Spin arrangement is considered at $H\approx H_c$ and $H<H_c$ in subsections \ref{trans2} and \ref{trans1}, respectively. The spin-wave spectrum is discussed in subsection \ref{swspec}.

\subsection{Transition from the collinear phase to the noncollinear one at $H=H_c$}
\label{trans2}

It it shown in the previous section that the spectrum \eqref{specres} of the collinear phase becomes unstable when $\Lambda<0$, i.e., when $H<H_c$. The instability takes place at momentum ${\bf k}_{c0}$ which absolute value is given by Eq.~\eqref{kc} and the direction is determined by small $1/S$-corrections. This instability signifies a transition to a noncollinear phase. We discuss now the spin arrangement in this phase using the approach of Bose-Einstein condensation of magnons. \cite{bat84} 

There are two equivalent minima in the bare spectrum \eqref{specres} at momenta $\pm{\bf k}_{c0}$. The instability of the spectrum at ${\bf k}=\pm{\bf k}_{c0}$ at $H$ slightly smaller than $H_c$ means a "condensation" of magnons at states with these momenta. It is shown below that at $H<H_c$ the energy can be lower by varying the value of the momentum ${\bf k}_c\|{\bf k}_{c0}$ at which the condensation takes place. Thus, $k_c$ depends on $H$ so that $k_c=k_{c0}$ at $H=H_c$ and $k_c<k_{c0}$ at $H<H_c$. Then we imply below that $k_c\ne k_{c0}$.

At $H<H_c$ one should replace operator $a_{\pm{\bf k}_c}$ and $a^\dagger_{\pm{\bf k}_c}$ as follows
\begin{equation}
\label{tr}
\begin{array}{l}
\displaystyle a_{\pm{\bf k}_c} \mapsto a_{\pm{\bf k}_c} + e^{i\varphi_1}\sqrt{\mathfrak N \rho_{{\bf k}_c}},\\
\displaystyle a^\dagger_{\pm{\bf k}_c} \mapsto a^\dagger_{\pm{\bf k}_c} + e^{-i\varphi_1}\sqrt{\mathfrak N \rho_{{\bf k}_c}},
\end{array}
\end{equation}
where $\rho_{{\bf k}_c}$ is the density of condensed particles and $\varphi_1$ is a phase constant which should be found from the demand that linear terms disappear in the Hamiltonian arising after transformation \eqref{tr} (see, e.g., Ref.~\cite{popov}). Taking into account the equivalence of minima at $+{\bf k}_c$ and $-{\bf k}_c$ we assume in Eq.~\eqref{tr} that $\rho_{{\bf k}_c} = \rho_{-{\bf k}_c}$ and similar equalities are implied below for harmonics of ${\bf k}_c$. One has from Eqs.~\eqref{h2} and \eqref{h4ls} for the term in the Hamiltonian linear in $a^\dagger_{{\bf k}_c}$
\begin{equation}
\label{linear1}
\sqrt{\mathfrak N \rho_{{\bf k}_c}}
\left(
e^{i\varphi_1}E_{{\bf k}_c} + e^{-i\varphi_1}B_{{\bf k}_c} + 3A\rho_{{\bf k}_c} 
\left(
e^{i\varphi_1} - \frac14 e^{i3\varphi_1} - \frac34 e^{-i\varphi_1}
\right)
\right)a^\dagger_{{\bf k}_c}.
\end{equation}
We obtain the term linear in $a_{{\bf k}_c}$ conjugating Eq.~\eqref{linear1}. In order expression \eqref{linear1} to be equal to zero at nonzero positive $\rho_{{\bf k}_c}$, $\varphi_1$ should be equal to $\pm\pi/2$. We choose plus sign below and one obtains that Eq.~\eqref{linear1} is equal to zero when
\begin{equation}
\label{linsol1}
\begin{array}{l}
\displaystyle \varphi_1 = \frac \pi2,\\
\displaystyle \rho_{{\bf k}_c} = - \frac{E_{{\bf k}_c} - B_{{\bf k}_c}}{6A} = -\frac{D(k_c-k_{c0})^2 + \Lambda}{6A}.
\end{array}
\end{equation}
It is seen that positive solution for $\rho_{{\bf k}_c}$ exists only if $\Lambda<0$, i.e., at $H<H_c$. Notice that the solution \eqref{linsol1} corresponds also to a result of minimization with respect to $\varphi_1$ and $\rho_{{\bf k}_c}$ of the correction to the energy arising after transformation \eqref{tr} and given by
\begin{equation}
\label{e1}
\frac{{\cal E}_1}{\mathfrak N} = 2\rho_{{\bf k}_c} (E_{{\bf k}_c} + B_{{\bf k}_c} \cos2\varphi_1) +  3A \rho_{{\bf k}_c}^2 (1-\cos2\varphi_1).
\end{equation}
Minimization of ${\cal E}_1$ with respect to $k_c$ gives $k_c=k_{c0}$. The deviation of $k_c$ from $k_{c0}$ arises after taking into account condensation of magnons at states characterized by harmonics of ${\bf k}_c$.

It is seen from Eq.~\eqref{h4ls} that transformation \eqref{tr} leads also to terms in the Hamiltonian linear in $a_{\pm3{\bf k}_c}$ and $a^\dagger_{\pm3{\bf k}_c}$. In order to cancel these terms one has to perform the following transformation of $a_{\pm3{\bf k}_c}$ and $a^\dagger_{\pm3{\bf k}_c}$ similar to \eqref{tr}
\begin{equation}
\label{tr3}
\begin{array}{l}
\displaystyle a_{\pm3{\bf k}_c} \mapsto a_{\pm3{\bf k}_c} + e^{i\varphi_3}\sqrt{\mathfrak N \rho_{3{\bf k}_c}},\\
\displaystyle a^\dagger_{\pm3{\bf k}_c} \mapsto a^\dagger_{\pm3{\bf k}_c} + e^{-i\varphi_3}\sqrt{\mathfrak N \rho_{3{\bf k}_c}}.
\end{array}
\end{equation}
As a result the term in the Hamiltonian linear in $a^\dagger_{3{\bf k}_c}$ arising after transformation \eqref{tr3} has the form
\begin{equation}
\label{linear3}
\sqrt{\mathfrak N}
\left(
\sqrt{\rho_{3{\bf k}_c}}
\left(
e^{i\varphi_3}E_{3{\bf k}_c} + e^{-i\varphi_3}B_{3{\bf k}_c} 
\right)
+ 
2iA\rho^{3/2}_{{\bf k}_c}
+ 
A\rho_{{\bf k}_c}\sqrt{\rho_{3{\bf k}_c}}
\left(
7e^{i\varphi_3} - 5e^{-i\varphi_3}
\right)
\right)a^\dagger_{3{\bf k}_c}.
\end{equation}
It is equal to zero if
\begin{equation}
\label{linsol3}
\begin{array}{l}
\displaystyle \varphi_3 = -\frac \pi2,\\
\displaystyle \sqrt{\rho_{3{\bf k}_c}} = -\sqrt{\rho_{{\bf k}_c}}
\frac13
\frac{D(k_c-k_{c0})^2 + \Lambda}{D(3k_c-k_{c0})^2 - 2D(k_c-k_{c0})^2 - \Lambda}.
\end{array}
\end{equation}
Solution \eqref{linsol3} corresponds also to the result of minimization with respect to $\varphi_3$ and $\rho_{3{\bf k}_c}$ of the correction to the energy arising after transformation \eqref{tr3} that has the form
\begin{equation}
\label{e31}
\frac{{\cal E}_3}{\mathfrak N} =  
2\rho_{3{\bf k}_c} (E_{3{\bf k}_c} + B_{3{\bf k}_c} \cos2\varphi_3)
+8A\rho_{{\bf k}_c}^{3/2}\sqrt{\rho_{3{\bf k}_c}}\sin\varphi_1\sin\varphi_3 
+2A(7-5\cos2\varphi_3)\rho_{{\bf k}_c}\rho_{3{\bf k}_c}.
\end{equation}
Condensation of magnons at states characterized by harmonics of ${\bf k}_c$ makes it possible to lower the energy further by varying the absolute value of ${\bf k}_c$ at any given $H$. To show this let us consider first terms in Eqs.~\eqref{e1} and \eqref{e31} which are the only terms depending on $k_c$ in these expressions. Their explicit forms are $2\rho_{{\bf k}_c}(D(k_c-k_{c0})^2 + \Lambda)$ and $2\rho_{3{\bf k}_c}(D(3k_c-k_{c0})^2 + \Lambda)$, respectively. It seen that at any fixed $\rho_{{\bf k}_c}$ and $\rho_{3{\bf k}_c}$ a small deviation of $k_c$ from $k_{c0}$ (to be precise, small decreasing) raise ${\cal E}_1$ and lower ${\cal E}_3$. The total energy lower in this case because ${\cal E}_1$ rises quadratically with $(k_c-k_{c0})$ and ${\cal E}_3$ lower linearly with $(k_c-k_{c0})$. As a result minimization of the total energy correction ${\cal E}_1+{\cal E}_3$ with respect to $k_c$ gives in the leading order in $\Lambda/Dk_{c0}^2$
\begin{equation}
\label{kccor}
k_c=k_{c0}\left( 1 - \frac{1}{24}\left(\frac{\Lambda}{Dk_{c0}^2}\right)^2 \right).
\end{equation}
Notice also that ${\bf k}_c$ should be parallel to ${\bf k}_{c0}$ because the absolute value of negative anisotropic corrections to $D$ discussed in the previous section have a maximum when ${\bf k}_c\|{\bf k}_{c0}$. It is seen from Eqs.~\eqref{linsol3} and \eqref{kccor} that $\rho_{3{\bf k}_c} \ll \rho_{{\bf k}_c}$ if $H\approx H_c$. We use this fact in Eqs.~\eqref{linear3} and \eqref{e31} discarding terms of powers in $\rho_{3{\bf k}_c}$ higher than $1/2$ and 1, respectively, and omitting terms with higher harmonics in Eqs.~\eqref{linear1}, \eqref{e1}, \eqref{linear3} and \eqref{e31}. 

It is easy to show that the phase factor of $(2n+1)$-th harmonic of ${\bf k}_c$ should be equal to $\pi/2$ or $-\pi/2$ in order the linear terms can vanish in the Hamiltonian. One has to take into account the fact that terms in the Hamiltonian linear in $a_{(2n+1){\bf k}_c}$ and $a_{(2n+1){\bf k}_c}^\dagger$ stem from ${\cal H}_{q}$ with even $q$ and that 
$
E_{(2n+1){\bf k}_c}\approx B_{(2n+1){\bf k}_c}\approx SA
$ 
if $n$ is not very large. Closer examination gives
\begin{equation}
\label{phi}
\varphi_{2n+1} = (-1)^n \frac\pi2.
\end{equation}
For instance, it is seen from Eq.~\eqref{e31} that $\varphi_1$ and $\varphi_3$ should have different signs to minimize the second term. The validity of Eq.~\eqref{phi} for higher harmonics is not so evident for large enough $|\Lambda|$, at which densities of condensed particles at states characterized by neighboring higher harmonics differ several times only. Meantime particular numerical calculations show the validity of Eq.~\eqref{phi} in this case too.

Terms in the Hamiltonian with odd number of operators $a$ and $a^\dagger$ which are proportional to Fourier components of the non-diagonal parts of the dipolar tensor $Q^{xz}$ (see, e.g., Eq.~\eqref{h3}) lead to condensation of magnons at states characterized by even harmonics of ${\bf k}_c$. It is easy to show that phase constants $\varphi_{2n}$ should be equal to zero modulo $\pi$ in corresponding transformations similar to \eqref{tr} and \eqref{tr3}. The density of condensed particles are very small at states characterized by even harmonics of ${\bf k}_c$. For instance, $\rho_{2{\bf k}_c} \propto \rho_{{\bf k}_c}^2 (Q^{xz}_{2{\bf k}_c}/A)^2 \ll \rho_{{\bf k}_c}$. Notice that $\rho_{2n{\bf k}_c}=0$ when ${\bf k}_c\perp {\bf H}$ because $Q^{xz}_{2n{\bf k}_c}=0$ in this case. The density of condensed particles in the state with zero momentum $\rho_{\bf 0}$ is equal to zero at any ${\bf k}_c$ direction because $Q^{xz}_{\bf 0}=0$.

\begin{figure}
\centering
\includegraphics[scale=0.9]{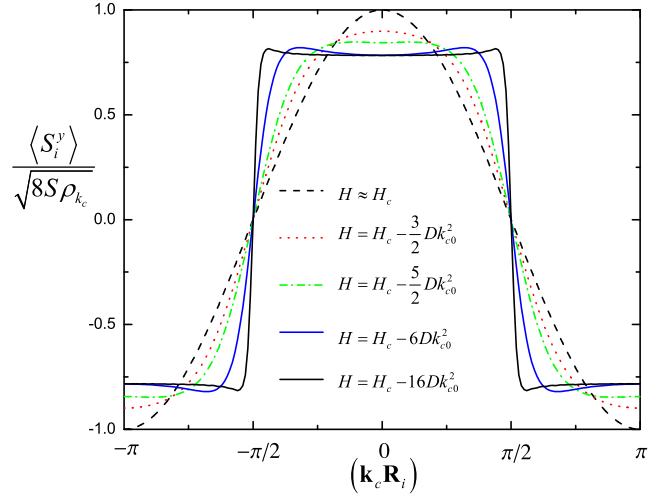}
\caption{(Color online). The value $\langle S_i^y \rangle/\sqrt{8S\rho_{{\bf k}_c}}$ given by Eq.~\eqref{si} is shown in the noncollinear phase (i.e., at $H<H_c$) to demonstrate the field evolution of the domain pattern profile forming by off-plane spins components (see Fig.~\ref{phases}). It is seen that the profile is sinusoidal at $H\approx H_c$ that evolves upon the field decreasing into a meander-like and then into a step-like form. Notice that the period of the stripe structure rises upon the field decreasing as it is illustrated in Fig.~\ref{kcfig}.
\label{stripe}} 
\end{figure}

One can find the spin ordering in the noncollinear phase at $H\approx H_c$ using Eqs.~\eqref{linsol1}, \eqref{linsol3} and \eqref{kccor}. We obtain mean values of the transverse spins components
$\langle S^x_{\bf k} \rangle \approx \sqrt{2S} \langle a_{\bf k} + a^\dagger_{-\bf k}\rangle/2$ and
$\langle S^y_{\bf k} \rangle \approx \sqrt{2S} \langle a_{\bf k} - a^\dagger_{-\bf k}\rangle/2i$
using the fact that $\langle a_{n{\bf k}_c}\rangle = e^{i\varphi_n}\sqrt{\mathfrak N \rho_{n{\bf k}_c}}$. Only even and odd harmonics contribute to $\langle S^x_i\rangle$ and $\langle S^y_i\rangle$, respectively, because $\varphi_{2n+1}=\pm\pi/2$ and $\varphi_{2n}=0\mod\pi$ and we have 
\begin{equation}
\label{si}
\langle S_i^y \rangle = \sqrt{8S\rho_{{\bf k}_c}} 
	\left(
	\cos({\bf k}_c{\bf R}_i) + \sum_{n=1}^\infty (-1)^n\sqrt{\frac{\rho_{(2n+1){\bf k}_c}}{\rho_{{\bf k}_c}}} \cos((2n+1){\bf k}_c{\bf R}_i)
	\right).
\end{equation}
Due to the smallness of densities of condensed particles at states with even harmonics we do not discuss $\langle S^x_i\rangle$ here. As $\rho_{(2n+1){\bf k}_c}\ll \rho_{{\bf k}_c}$ when $H\approx H_c$, only the first few terms play in Eq.~\eqref{si} at such $H$. We plot the value $\langle S_i^y \rangle/\sqrt{8S\rho_{{\bf k}_c}}$ in Fig.~\ref{stripe} for $H\approx H_c$ and $H=H_c - 1.5Dk_{c0}^2$ using Eqs.~\eqref{linsol1}, \eqref{linsol3}, \eqref{kccor}, and \eqref{si}. It is seen that $\langle S_i^y \rangle$ has a sinusoidal profile at $H\approx H_c$ as it is shown in Fig.~\ref{phases} that is slightly distorted at smaller field due to the higher harmonics in Eq.~\eqref{si}. Higher harmonics in Eq.~\eqref{si} become important at $H<H_c - 1.5Dk_{c0}^2$ when they change considerably the profile of $\langle S_i^y \rangle$ (i.e., the domain pattern profile) which we discuss in the next subsection. 

Let us compare energies of the stripe phase and the collinear canted phase at $H<H_c$. The last one can be found performing the transformation for operators $a_{\bf 0}$ and $a^\dagger_{\bf 0}$ similar to \eqref{tr} and \eqref{tr3} and minimizing the corresponding expression for the energy correction. The result is
\begin{equation}
\label{e0}
\frac{{\cal E}_0}{\mathfrak N} = 
-\frac{(E_{\bf 0} - B_{\bf 0})^2}{4A} = 
-\frac{\left(Dk_{c0}^2 - |\Lambda|\right)^2}{4A}.
\end{equation}
This expression gives the value of the energy correction higher than that of the stripe phase. For example, corresponding values for $H=H_c'$ are $-0.179\Lambda^2/A$ and $-0.063\Lambda^2/A$ for the stripe and the canted collinear phase, respectively, where $H_c'$ given by Eq.~\eqref{hc1sw} is the field below which the classical spin-wave spectrum of the canted phase is stable. The difference between energies of the collinear canted phase and the stripe one decreases as the field reduces.

\subsection{$H<H_c$}
\label{trans1}

We demonstrate in Appendix~\ref{enc} that expression for the energy of the noncollinear phase is much more complicated at $H<H_c$ because one has to take into account corrections to the energy containing $\rho_{(2n+1){\bf k}_c}$ with $n>1$ for accurate finding of the stripe profile and $k_c$. The corresponding values of $\rho_{(2n+1){\bf k}_c}$ and $k_c$ have to be found by numerical minimization of the energy. The number of harmonics of ${\bf k}_c$ to be taken into account in particular calculations rises as the field decreases. For example, we find that one can use only five harmonics for $H=H_c-2.5Dk_{c0}^2$ whereas their number should be increased to 51 for $H=H_c-16Dk_{c0}^2$. This circumstance leads to a restriction on the field interval which can be considered practically. The smallest field value we have managed to reach is $H=H_c-16Dk_{c0}^2$.

We plot $\langle S_i^y \rangle/\sqrt{8S\rho_{{\bf k}_c}}$ in Fig.~\ref{stripe} for some $H$ values using Eq.~\eqref{si} and results of numerical calculation of $\rho_{(2n+1){\bf k}_c}$. We used Eqs.~\eqref{linsol1}, \eqref{linsol3}, \eqref{kccor} for $H=H_c-1.5Dk_{c0}^2$ and Eqs.~\eqref{rhohc1} for $H=H_c-2.5Dk_{c0}^2$. It is seen that the sinusoidal profile evolves upon the field decreasing into a meander-like and then into a step-like form. Then, the domain wall density decreases from unity to a value much smaller than unity upon the field decreasing.

\begin{figure}
\centering
\includegraphics[scale=0.9]{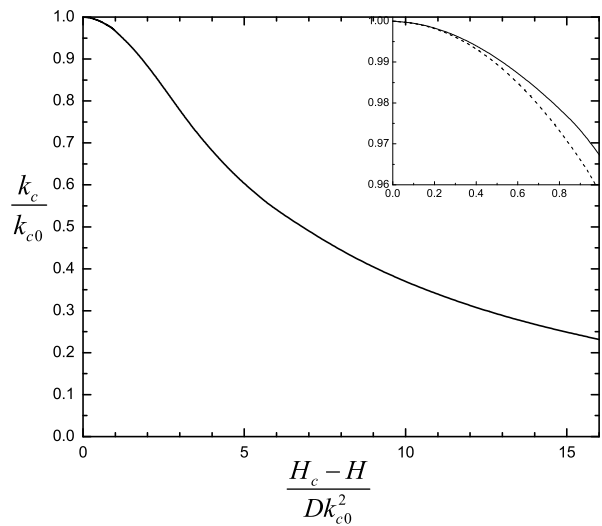}
\caption{Dependence of the absolute value of the momentum ${\bf k}_c$ describing the stripe structure on the magnetic field found by numerical energy minimization taking into account 51 harmonics of ${\bf k}_c$. The inset shows the neighborhood of point $H=H_c$, where the dependence of $k_c$ on $H$ is described by approximate analytic expression \eqref{kccor} (dashed line). 
\label{kcfig}} 
\end{figure}

Dependence of $k_c$ on $H$ is illustrated in Fig.~\ref{kcfig}. Notice that the period of the stripe structure given by $2\pi/k_c$ rises very fast upon the field decreasing. This finding is in qualitative agreement with results of Ref.~\cite{stripes}, where a very large period of the stripe pattern was found at $H=0$ when $A\gg\omega_0$.

\subsection{Spin-wave spectrum in the noncollinear phase}
\label{swspec}

There are two types of corrections to the bilinear part of the Hamiltonian \eqref{h2} arising after transformation \eqref{tr}. The first one leads to the following simple renormalization of $E_{\bf k}$ and $B_{\bf k}$:
\begin{equation}
\label{shift}
E_{\bf k} \mapsto E_{\bf k} + 7A\rho_{{\bf k}_c}, \quad B_{\bf k} \mapsto B_{\bf k} - 5A\rho_{{\bf k}_c}.
\end{equation}
This renormalization gives rise to an additional term equal to $2|\Lambda|$ in the second bracket in Eq.~\eqref{specres} and the spectrum takes the form
\begin{equation}
\label{specresnc}
\epsilon_{\bf k}^2 \approx \left(Dk^2 + H \right)\left(D(k-k_c)^2 + |\Lambda|\right).
\end{equation}
Higher harmonics contribute to renormalization of $E_{\bf k}$ and $B_{\bf k}$ as well but their effect is small at $|\Lambda|< Dk_{c0}^2$ because corresponding densities are much smaller than $\rho_{{\bf k}_c}$.

Another sort of corrections to the bilinear part of the Hamiltonian include the so-called umklapp interaction having the form
\begin{equation}
\label{h2u}
A\rho_{{\bf k}_c} \sum_{\bf k} 
\left[
\frac72 
a^\dagger_{\bf k} a_{{\bf k} \pm 2{\bf k}_c}
- 
\frac54 
\left(
a_{\bf k}a_{-{\bf k} \pm 2{\bf k}_c} + 
a^\dagger_{\bf k}a^\dagger_{-{\bf k} \pm 2{\bf k}_c}
\right)
\right].
\end{equation}
There are also terms given by Eq.~\eqref{h2u} with $2n{\bf k}_c$ and $\rho_{n{\bf k}_c}$ instead of $2{\bf k}_c$ and $\rho_{{\bf k}_c}$ with $n>1$ which can be neglected at $|\Lambda|< Dk_{c0}^2$. The umklapp interaction makes much more complicated the spin-wave spectrum analysis (see, e.g., Ref.~\cite{maleyev} for some detail). Meantime it can be shown that corrections to the spectrum square from umklapp terms \eqref{h2u} are negligible for momenta which absolute values are not very close to $k_c$: $D(k-k_c)^2\gtrsim |\Lambda|$. Then the spectrum is given by Eq.~\eqref{specresnc} in this case. Carrying out of much more complicated analysis for $k\approx k_c$ is out of the scope of the present work.

\section{Conclusion}
\label{con}

In conclusion, we discuss 2D Heisenberg ferromagnet with dipolar forces and out-of-plane easy-axis one-ion anisotropy in in-plane magnetic field described by the Hamiltonian \eqref{ham0}. We consider the classical spin-wave spectrum assuming a collinear spin arrangement and find in accordance with previous results \cite{emills} that the spectrum is unstable in a range of the field $H_c'<H<H_c$, where $H_c$ is the saturation field given by Eq.~\eqref{hc2t0} and $H_c^\prime$ is given by Eq.~\eqref{hc1sw}. The instability arises at an incommensurate momentum ${\bf k}_{c0}$ which direction is not fixed in the linear spin-wave approximation and which absolute value is given by Eq.~\eqref{kc}. Thus, we conclude following Ref.~\cite{emills} that the collinear spin arrangement is unstable at $H<H_c$ with respect to a stripe domain formation with a large period (see Fig.~\ref{phases}).

To discuss properties of the non-collinear phase we consider first $1/S$ corrections to the spectrum of the collinear phase at $H>H_c$ and find that they fix the direction of ${\bf k}_{c0}$ so that it depends on the field direction as it is shown in Fig.~\ref{angle}. We discuss the noncollinear phase using the technique of the Bose-Einstein condensation of magnons suggested in Ref.~\cite{bat84}. Appearance of the stripe spin arrangement corresponds in this technique to a "condensation" of magnons at $H<H_c$ at states characterized by a momentum ${\bf k}_c\|{\bf k}_{c0}$ and its harmonics (i.e., $n{\bf k}_c$, where $n$ is integer). It is the unusual result of the present analysis that the vector ${\bf k}_c$ varies with the field (remaining parallel to ${\bf k}_{c0}$) so that $k_c=k_{c0}$ at $H=H_c$ and $k_c<k_{c0}$ at $H<H_c$. The origin of this behavior is in condensation of magnons at states characterized by harmonics of ${\bf k}_c$.

We demonstrate that the energy of the stripe phase is lower than the energy of any collinear spin arrangement at $H<H_c$ but the difference between these energies decreases as the field reduces. We have not found any sign of transition to a collinear phase at $H\lesssim H_c$. Then, $H_c^\prime$ given by Eq.~\eqref{hc1sw} is overestimated in the previous work \cite{emills} in which it was obtained from the condition of the spin-wave spectrum stability in the collinear canted phase. It should be noted that this our finding is in qualitative agreement with results of Ref.~\cite{stripes}, where it was shown, in particular, that at $H=0$ the difference between energies of the stripe pattern and the collinear phase is very small in the limiting case of $A\gg\omega_0$ discussed in the present paper.

Expression for values of spins components perpendicular to the film forming the domain pattern is given by Eq.~\eqref{si}. Coefficients $\rho_{n{\bf k}_c}$ and the value of $k_c$ are obtained by numerical minimization of the ground state energy. The transformation of the domain pattern upon the variation of the field is shown in Fig.~\ref{stripe}. It is seen that the domain pattern has a sinusoid profile at $H\approx H_c$ that transforms to a step-like profile upon the field decreasing so that the domain wall density decreases from unity to a value much smaller than unity. The dependence of $k_c$ on the field is shown in Fig.~\ref{kcfig}. It is seen that the period of the stripe pattern ($2\pi/k_c$) rises very fast as the field decreases. This finding is also in qualitative agreement with results of Ref.~\cite{stripes}, where it was shown that at $H=0$ the period of the stripe structure is very large at $A\gg\omega_0$.

We find that the spin-wave spectrum in the noncollinear phase is given by Eq.~\eqref{specresnc} for momenta $D(k-k_c)^2\gtrsim |\Lambda|$ if $|\Lambda|< Dk_{c0}^2$.

\begin{acknowledgments}

This work was supported by Russian Science Support Foundation, President of Russian Federation (grant MK-1056.2008.2), RFBR grants 09-02-00229 and 07-02-01318, and Russian Programs "Quantum Macrophysics", "Strongly correlated electrons in semiconductors, metals, superconductors and magnetic materials" and "Neutron Research of Solids".

\end{acknowledgments}

\appendix

\section{Energy of the noncollinear phase}
\label{enc}

We present in this appendix some detail of the noncollinear phase energy calculation that was used in finding the value of $k_c$ drawn in Fig.~\ref{kcfig} and the domain pattern profile presented in Fig.~\ref{stripe}. It is convenient to express the correction to the energy arising after transformations \eqref{tr} and \eqref{tr3} (and similar ones for higher harmonics of ${\bf k}_c$) in the following form: 
\begin{equation}
\label{en}
	{\cal E} = \sum_{n=0}^{\infty}{\cal E}_{2n+1},
\end{equation}
where ${\cal E}_{2n+1}$ contains all terms depending on $\rho_{(2n+1){\bf k}_c}$ and not depending on $\rho_{(2i+1){\bf k}_c}$ with $i>n$. In particular, one has for the first three terms in Eq.~\eqref{en}
\begin{eqnarray}
\label{e11}
{\cal E}_1 &=& 2\rho_{{\bf k}_c}\left(D(k_c-k_{c0})^2-|\Lambda|\right) + 6A\rho_{{\bf k}_c}^2,\\
\label{e3}
{\cal E}_3 &=&  
-8A\rho_{{\bf k}_c}^{3/2}\sqrt{\rho_{3{\bf k}_c}} 
+24A\rho_{{\bf k}_c}\rho_{3{\bf k}_c}
+6A\rho_{3{\bf k}_c}^2
+2\rho_{3{\bf k}_c}\left(D( 3k_c-k_{c0})^2 - |\Lambda| \right),\\
\label{e5}
{\cal E}_5 &=&  
-24A\rho_{{\bf k}_c}\sqrt{\rho_{3{\bf k}_c}\rho_{5{\bf k}_c}} 
+24A\rho_{3{\bf k}_c}\sqrt{\rho_{{\bf k}_c}\rho_{5{\bf k}_c}} 
+24A(\rho_{{\bf k}_c} + \rho_{3{\bf k}_c})\rho_{5{\bf k}_c}
+6A\rho_{5{\bf k}_c}^2\nonumber\\
&&{}+2\rho_{5{\bf k}_c}\left(D( 5k_c-k_{c0})^2 - |\Lambda| \right),
\end{eqnarray}
where we take into account Eq.~\eqref{phi} for phase factors $\varphi_{2n+1}$. We can restrict ourself by some first terms in Eq.~\eqref{en} to perform the numerical minimization of $\cal E$ because the value of $\rho_{(2n+1){\bf k}_c}$ decreases as $n$ rises. Meantime, the number of terms to be taken into account in Eq.~\eqref{en} rises as the field decreases. For example, we find that one can use only terms \eqref{e11}--\eqref{e5} for $H=H_c-2.5Dk_{c0}^2$. The corresponding values of $k_c$ and densities are given by
\begin{eqnarray}
k_c & \approx & 0.832 k_{c0},\\
\label{rhohc1}
\rho_{{\bf k}_c} & \approx & 1.13 \frac{|\Lambda|}{6A},
\quad
\sqrt{\frac{\rho_{3{\bf k}_c}}{\rho_{{\bf k}_c}}} \approx 0.183,
\quad
\sqrt{\frac{\rho_{5{\bf k}_c}}{\rho_{{\bf k}_c}}} \approx 0.0325,
\quad
\sqrt{\frac{\rho_{7{\bf k}_c}}{\rho_{{\bf k}_c}}} \approx 0.0057,
\end{eqnarray}
which, in particular, have been put into Eq.~\eqref{si} to plot $\langle S_i^y \rangle/\sqrt{8S\rho_{{\bf k}_c}}$ in Fig.~\ref{stripe}. We have obtained that one has to take into account 26 terms in Eq.~\eqref{en} for $H=H_c-16Dk_{c0}^2$.

\section*{References}

\bibliography{dipfm}

\end{document}